# Variation Aware Training of Hybrid Precision Neural Networks with 28nm HKMG FeFET Based Synaptic Core

**Sunanda Thunder and Po-Tsang Huang**
[1] International College of Semiconductor Technology, National Yang-Ming Chiao Tung University, Hsinchu, Taiwan

**ABSTRACT** The plethora of data generated by edge devices and IoT devices has made machine learning the default choice of everyone for solving many complex tasks. Deep neural networks (DNNs) while deployed to perform several tasks such as speech and image recognition have shown superior performance compared to trivial machine learning algorithms. The applications like intelligent healthcare monitoring systems, smartwatches, or automatic cars require real-time processing of the data or image, which is done by machine learning algorithms with higher efficiency than humans. There are two possible methods for artificial intelligence-

1. non-Von-Neumann hardware-based implementation of neural network.

2. Traditional computer science base approach for neural networks or traditional Von-Neumann architecture-based implementation of neural networks.

The standard Von-Neumann performance of neural networks, where the memory and the computation parts are segregated, severely suffers from latency with the rising number of edge devices. However, the plethora of usage of edge devices in our daily life foists stringent restrictions on latency, device area, and power consumption for the hardware. Therefore, we need to take the route beyond CMOS-based mixed-signal implementation of neural networks. However, the training of emerging non-volatile memory (eNVM) based DNNs is rigorous, which necessitates the search for a novel computing framework for this application. A crossbar array with eNVM devices calculates the huge, weighted vector-matrix computation after storing the weight of the synapse as conductance states. However, the main challenge is the weight update process in a reliable manner in presence of device variations. This work proposes a hybrid-precision neural network training framework with an eNVM based computational memory unit executing the weighted sum operation and another SRAM unit, which stores the error in weight update during backpropagation and the required number of pulses to update the weights in the hardware. The hybrid training algorithm for MLP based neural network with 28 nm ferroelectric FET (FeFET) as synaptic devices achieves inference accuracy up to 95% in presence of device and cycle variations. The architecture is primarily evaluated using behavioral or macro-model of FeFET devices with experimentally calibrated device variations and we have achieved accuracies compared to floating-point implementations.

## I. INTRODUCTION

In the recent era, *artificial intelligence* has proved to be a *de-facto* choice of many computer scientists and researchers for solving complex problems, and henceforth the effort of making our computers more intelligent by soliciting the "Artificial Intelligence" has seen a fresh movement [1]. Although in the past machine learning has been used extensively to solve many difficult problems, it is difficult to express many problems by mathematical formalism, which necessitates infusing learning ability in the machines through the neural implementation of the network. The neural network is a mapping of one set of functions to another set through synaptic weights and activation function, where one set belongs to the input and the other set belongs to the output. The quintessential representation of a simple neural network is a multi-level perceptron (MLP) neural network, which will be the focus of this research. The learning process of the neural network involves obtaining the proper values for the synaptic weights using a known dataset. So that it can perform any task on an unknown dataset [2,3]. The training process can be divided into three different stages.

1. Forward propagation
2. Back-Propagation.
3. Weight update

The input data is propagated to the neural network in a forward direction during the first step. The output is calculated based on the data propagated in the forward direction in the initial iteration. This step is followed by a calculation of the error at the final output layer by comparing the output with the desired result. During backpropagation, this error value is propagated in the backward direction from the output layer towards the input layer and the gradient of the cost function for all synaptic weights is calculated. This gradient of the cost function is used for updating the synaptic weights in the final stage of operation. The cost function is minimized but iteratively repeats this whole process for the entire data set, which makes this training an exhaustive process.

In the past few years, the world has seen a rise in the utilization of edge devices, especially in health care systems such as a smartwatch or smart band. The amount of real-time data processed by these devices has increased the need for data-centric computing. In traditional Von-Neumann architecture, the segregation of computing units and memory units creates the biggest bottleneck in implementing efficient data-centric computing, introducing high latency in data transfer, while the volatile nature of memory units also increases the energy consumption. Although recent development of graphical processing units (GPUs) has accelerated neural network training to some extent, the power-hungry nature of these GPUs makes them on-ideal for the training of large neural networks. This motivates the investigation of emerging non-volatile memories (eNVMs) for computing-in-memory (CIM) purposes [4-13]. The core of this CIM architecture is the memory array, which is usually designed by the eNVMs. This memory array is used to perform vector-matrix multiplication, which is done by exploiting *Kirchhoff's Current Law* (KCL). Quintessentially these NVM-based arrays can represent the synapse between two layers of fully connected neurons, where the conductance of the eNVM devices are mapped to the synaptic weights by one-to-one function. The eNVMs for neuromorphic applications have some specific requirements from the device's perspective, which are essentially CMOS compatibility, scalability, multi-level programming, high endurance, low variation, and low power consumption. Among many eNVMs hafnium oxides ($HfO_2$) based ferroelectric memory turns out to be one of the most promising ones. This feature can be attributed to the CMOS compatibility, high endurance, higher ON-state current to OFF-state current ratio, and multi-level operation. Recent progress in the research on 28nm high-k metal gate (HKMG) based ferroelectric field-effect transistors (Fe-FETs), and deeply scaled ferroelectric fin field-effect transistors (Fe-finFETs) corroborates this fact. This progress has accelerated the application of deeply scaled FeFETs for computing-in-memory (CIM) applications [14-22]. However, device-to-device (D2D) variations in $HfO_2$ based deeply scaled FeFETs pose a severe threat towards accurately executing vector-matrix multiplication operations. Despite numerous efforts to improve the D2D variations in FeFETs, the random distribution of ferroelectric grains in deeply scaled FeFETs poses a serious threat towards its large-scale memory array level applications, especially for neuromorphic applications [23-30]. This research aims towards eliminating the impact of device variation in FeFETs on their application towards neuromorphic application, especially during training of deep neural networks (DNN) by deploying a hybrid precision training algorithm.

We commenced our work with 28nm high-k metal gate (HKMG) based ferroelectric FETs, fabricated in GlobalFoundries. Devices with various dimensions were characterized and statistically modelled. We observed that as the devices are scaled below ~ 0.02 μm$^2$, device variation becomes severe and the program-erase operations lose their fidelity, especially for multilevel operations. The channel conductance ($G_{ch}$) in FeFETs is based on the arrangement of the dipoles in the high-k or ferroelectric layer of the gate stack. The increased stochasticity in grain distribution with scaling and polymorphism of $HfO_2$ based ferroelectric layers leads to a significant reduction in training accuracy compared to the software baseline. Several research has been conducted and methods like multiple devices per synapse, mixed-precision architecture, and accumulate weight update mechanism in 3T-1C have been proposed to increase the precision of the synaptic devices. However, most of these mechanisms increase the complexity of the neuromorphic system and increase the footprint of the chip. Although typically higher precision of weights yields high training accuracy, it is shown by [26] that it is possible to achieve high accuracy even with binary precision for forwarding and backpropagation, keeping high precision for the gradient. A recent study on mixed-precision training by [26], is also stemmed from this idea. However, that very implementation is limited to online training. Typically, online training of the neural network requires high WRITE-endurance and online training wears out the synaptic devices. This work mostly focuses on variation-aware hybrid precision training of FeFET based synaptic core. The FeFET based synapses are binary for forwarding and backpropagation. We have used additional static-random-access memory (SRAM) based memories along with the FeFET based synaptic core to store gradient values with high precision. The training starts with programming the complete network with all "0" and all "1" values respectively 100 times. This was followed by calculating the error in $G_{ch}$. The error statistics across devices and cycles were modeled using gaussian distribution and an invertible macro model for FeFET weight update was built. Once the device data and the statistics were accumulated the training of the neural network began. The gradient values were stored in a high-precision accumulator which was emulated by SRAM cells in the framework. The weight update was bitwise and was done only when the gradient value crossed the threshold limit. Finally, the weight to pulse value was calculated for each FeFETs and corresponding weights were programmed in the FeFET cells. The classification accuracy after the final weight update reached 98%, which is quite like the accuracy achieved by floating-point precision.

Cycles were modelled using gaussian distribution and an invertible macro model for FeFET weight update was built. Once the device data and the statistics were accumulated the training of the neural network began. The gradient values were stored in a high-precision accumulator which was emulated by SRAM cells in the framework. The weights update was bitwise and was done only when the gradient value crossed the threshold limit. Finally, the weight to pulse value was calculated for each FeFETs and corresponding weights were programmed in the FeFET cells. The classification accuracy after the final weight update reached 98%, which is quite like the accuracy achieved by floating-point precision.

## II. Conclusion

This work demonstrated the hybrid precision training of neural networks via experiment and simulations. Experimentally calibrated data was used for neural network simulation. The training algorithm demonstrated excellent immunity towards device variation and low precision data processing units. Deploying the hybrid-precision algorithm yielded inference accuracy above 95% for MNIST handwritten datasets with binary precision in presence of device variations.

## ACKNOWLEDGMENT

We are grateful to the Fraunhofer IPMS and GlobalFoundries for nanofabrication facilities and services.